\documentstyle[multicol,epsf,aps]{revtex}
\begin{document}
\title{Fragmentation with a Steady Source}
\author{E.~Ben-Naim$^1$ and P.~L.~Krapivsky$^{2,3}$}
\address{$^1$Theoretical Division and Center for Nonlinear Studies, 
 Los Alamos National Laboratory, Los Alamos, NM 87545}
\address{$^2$Center for Polymer Studies and Department of Physics, 
Boston University, Boston, MA 02215} 
\address{$^3$CNRS, IRSAMC, Laboratoire de Physique Quantique,
Universite' Paul Sabatier, 31062 Toulouse, France}
\maketitle 
\begin{abstract}
  We investigate fragmentation processes with a steady input of
  fragments.  We find that the size distribution approaches a
  stationary form which exhibits a power law divergence in the small
  size limit, $P_\infty(x)\sim x^{-3}$. This algebraic behavior is
  robust as it is independent of the details of the input as well as
  the spatial dimension.  The full time dependent behavior is obtained
  analytically for arbitrary inputs, and is found to exhibit a
  universal scaling behavior.

\end{abstract}
    
\begin{multicols}{2}
  
  Fragmentation underlies numerous natural
  phenomena\cite{har,shr,hab,red,ms}.  The quantity being ``split''
  can be the mass, momentum, or the area, and typically, fragments
  continue splitting independently of each other.  Examples include
  polymer degradation \cite{zm1}, energy cascades in
  turbulence\cite{man,bppv}, breakup of liquid droplets\cite{liq} and
  atomic nuclei\cite{nuc}, martensitic transformations\cite{rss,bk},
  shattering of solid objects\cite{im,odb}, and meteor impacts.
  Fragmentation also arises in several topics of computer
  science\cite{kgb,mahm,km}.

The simplest fragmentation models assume that the rate by which
fragments are produced is a function of their size
only\cite{char,zm2,fil,cr,plk}. In this study, we focus on the classic
``random scission model'' where the cutting is uniform and hence a
fragment is cut with a rate proportional to its size. In particular,
we are interested in situations where the system is subject to a
steady input of fragments. Such ``open'' systems may be realized
physically in impact fragmentation of solid objects where it is
possible to constantly supply the system with unfragmented objects.
  
Fragmentation in open systems has received much less attention than
fragmentation in closed systems.  We will show, however, that
fragmentation with input is actually {\em simpler} than the classical
counterpart as the system reaches a stationary state which is
remarkably robust.  Specifically, fragmentation with a steady source
is characterized by an algebraic divergence of the size distribution
in the small size limit, and this behavior is independent of the
particular form of the input.  Additionally, the time dependent
behavior, obtained analytically for arbitrary inputs, follows a
scaling behavior. These two features are shown to be closely related.

We start with a one-dimensional fragmentation process subject to
constant input of segments.  Here ``one-dimensional'' means that the
fragments are characterized by a single variable which we shall call
``length'' (the fragments can be viewed as segments).  Let the system
be initially empty and intervals whose length is within the range $(x,
x+dx)$ are added with rate $f(x)\,dx$.  Additionally, intervals are
cut with a constant spatially homogeneous rate; we set this rate equal
to unity without loss of generality.  Fragmentation with input has a
natural geometric interpretation.  Consider the segments as part of an
infinite line. The fragmentation process is equivalent to deposition
of point ``cracks'' on the line.  The line is initially ``immune'' to
fragmentation, but then segments of length $x$ become ``susceptible''
to fragmentation with rate $f(x)$. Hence, fragmentation with input is
equivalent to inhomogeneous fragmentation of a growing line.

The density $P(x,t)$ of intervals of length $x$ at time $t$ evolves
according to the following rate equation
\begin{equation}
\label{rate} 
{\partial P(x,t)\over \partial t}=-xP(x,t)+2\int_x^\infty dy\,P(y,t)+f(x).
\end{equation} 
The negative term on the right-hand side accounts for loss due to
fragmentation with the rate equal to the fragment size since the
cutting is uniform.  The gain term gives the increase in fragments of
size $x$ due to cutting of longer fragments. The last term accounts
for input of fragments of size $x$.

The size distribution can be determined by applying the Mellin
transformation.  The Mellin transform (or moment) of the
distribution, $M(s,t)=\int dx\, x^{s-1} P(x,t)$, satisfies 
\begin{equation}
\label{mel}
{\partial M(s,t)\over \partial t}={2-s\over s}M(s+1,t)+\hat f(s), 
\end{equation} 
where $\hat f(s)=\int dx\, x^{s-1} f(x)$ is the Mellin transform of
the input density $f(x)$. Although this hierarchy of equations is
infinite, its linear nature makes it tractable, as will be seen below.

We first examine what happens when $t\to\infty$.  In this limit, the
length density should approach the stationary distribution, $P(x,t)\to
P_\infty(x)$.  Setting the time derivative in Eq.~(\ref{mel}) to zero
gives the corresponding transform \hbox{$M_\infty(s)=\left(1+{2\over
s-3}\right)\hat f(s-1)$}.  Note that $\hat f(s+n)$ is the Mellin
transform of $x^{n}f(x)$, and $(s-m)^{-1}\hat f(s)$ is the transform
of $x^{-m}\int_x dy\, y^{m-1}f(y)$.  These two facts allow to perform
the inverse Mellin transform and yield $P_\infty(x)$ explicitly in
terms of the input function
\begin{equation}
\label{pinf} 
P_\infty(x)=x^{-1}f(x)+2\,x^{-3}\int_x^\infty dy\,yf(y).
\end{equation}
In the small size limit, the integral on the right-hand side of
Eq.~(\ref{pinf}) approaches the average length added per unit time,
$\lambda=\hat f(2)=\int dx\, x f(x)$. Thus, the length density becomes
purely algebraic
\begin{equation}
\label{alg}
P_\infty(x)\to 2\lambda x^{-3},\quad {\rm when}\quad x\to 0. 
\end{equation}
This behavior is robust as the first term on the right-hand side of
Eq.~(\ref{pinf}) always diverges slower than $x^{-3}$ in the limit
$x\to 0$ (otherwise, the total length input rate, $\int dx\,x f(x)$,
would be infinite). For a class of input densities, the
algebraic behavior may not be necessarily restricted to small sizes.
For example, for monodisperse inputs $f(x)=\lambda\delta(x-1)$, the
algebraic behavior extends to all sizes $x<1$, $P_\infty(x)=\lambda
x^{-1}\delta(x-1)+2\lambda x^{-3}$.

The general algebraic behavior should be contrasted with the
exponential length distribution found generally in the absence of
input.  Algebraic distributions have been observed
experimentally in fragmentation of solid objects such as rods,
spheres, and bricks \cite{im,odb}. Although the corresponding
exponents measured in these experiments are significantly lower,
typically between 1 and 2, it is worth noting that a steady source of
fragments can serve as a mechanism for generating algebraic
distributions.  Curiously, algebraic distributions with an exponent
close to $3$ were reported recently in social systems which can be
viewed as open ones (distributions of citations, of the number of
links to sites on the internet, etc.; see
e.g. Refs.\cite{ls,sr,barabasi}).

The limiting size distribution is ultimately related to the time
dependent behavior. This can be demonstrated using the following
heuristic argument. From Eq.~(\ref{mel}), the total length
$L(t)=M(2,t)$ grows linearly with time $\dot L(t)=\lambda$, and hence,
$L(t)=\lambda t$.  Similarly, the total number of fragments,
$N(t)=M(1,t)$ satisfies $\dot N(t)=\lambda t+\mu$, where $\mu=\hat
f(1)=\int dx f(x)$ is the number of segments added per unit time, and
consequently, $N(t)={1\over 2}\lambda t^2+\mu t$. These two time
dependent results imply that the average fragment length, $\langle
x\rangle =L/N$, decreases with time according to $\langle x\rangle\sim
t^{-1}$. For the length distribution to follow a scaling form, the
corresponding scaling variable must be $x/\langle x\rangle$.  The
prefactor is fixed by the total number of fragments, $N(t)\sim \lambda
t^2$, so the scaling form reads
\begin{equation}
\label{sclfrm}
P(x,t)\simeq \lambda t^3 F(xt).
\end{equation} 
This scaling form would be consistent with a time independent limiting
distribution only when $F(\xi)\sim \xi^{-3}$, thereby implying the
algebraic divergence (\ref{alg}).  

The full time dependent solution can be found using the Charlesby
method \cite{char}.  This method starts with a formal expansion of
the Mellin transform,
\begin{equation} 
\label{mst}
M(s,t)=\sum_{n=0}^\infty {t^n\over n!}\,M_n(s), 
\end{equation} 
and proceeds by solving for the functions $M_n(s)$ iteratively. Indeed,
substituting the expansion (\ref{mst}) into Eq.~(\ref{mel}) and
equating similar powers of time yields $M_0(s)=0$, $M_1(s)=\hat f(s)$
and $M_{n+1}(s)=-{s-2\over s}M_n(s+1)$ for $n\geq 2$.  Solving this
set of equations recursively gives
\begin{displaymath}
M_{n+1}(s)=(-1)^{n}{(s-1)(s-2)\over (s+n-1)(s+n-2)}\hat f(s+n).
\end{displaymath} 
To take advantage of the inversion rules used to obtain
Eq.~(\ref{pinf}), we re-write $M_{n+1}(s)$ as
\begin{displaymath}
M_{n+1}(s)=(-1)^{n}\left[1-{n(n+1)\over s+n-1}
+{n(n-1)\over s+n-2}\right]\hat f(s+n).
\end{displaymath} 
{}From Eq.~(\ref{mst}), the size distribution can be written as 
a power series 
\begin{equation}
\label{pxts}
P(x,t)=\sum_{n=0}^\infty {t^{n+1}(-x)^{n}\over (n+1)!}P_n(x),
\end{equation} 
where the inverse transform of $M_{n+1}(s)$ has been conveniently
written as $(-x)^{n}P_n(x)$.  The three terms in the above expression
for $M_{n+1}(s)$ can be inverted using the rules outlined before
Eq.~(\ref{pinf}). The final expression for $P_n(x)$ reads
\begin{equation}
\label{pnx}
P_n(x)=f_1(x)+{n(n+1)\over x}f_2(x)+{n(n-1)\over x^2}f_3(x),
\end{equation} 
with
\begin{eqnarray}
\label{fk}
f_1(x)&=&f(x),\nonumber\\
f_2(x)&=&\int_x^\infty dy\,f(y),\\
f_3(x)&=&\int_x^\infty dy\,y f(y).\nonumber
\end{eqnarray} 
Summing the three terms separately gives the fragment size
distribution 
\begin{equation}
\label{full}
P(x,t)=\sum_{k=1}^3 t^{k}f_k(x)F_k(xt),
\end{equation} 
with the scaling functions 
\begin{eqnarray}
\label{Fk}
F_1(z)&=&z^{-1}(1-e^{-z}),\nonumber\\ 
F_2(z)&=&e^{-z}, \\
F_3(z)&=&z^{-3}\left[2-(2+2z+z^2)e^{-z}\right].\nonumber
\end{eqnarray}
The function $F_3(z)$ has been obtained from the power series
\hbox{$F_3(z)=\sum_{n\geq 0}{(-z)^n\over n!(n+3)}$}.  One can verify
that the previous results for $N(t)$, $L(t)$, and
$P_{\infty}(x)$ agree with this solution.  Thus, we have obtained the
full time dependent solution for an {\em arbitrary} time independent
input $f(x)$.  

The size distribution of Eq.~(\ref{full}) exhibits scaling. Indeed, in
the limit $t\to\infty$, $x\to 0$ with the scaling variable $z=xt$ kept
finite, the third term in the sum on the right-hand side of
Eq.~(\ref{full}) dominates, and the anticipated scaling behavior of
Eq.~(\ref{sclfrm}) is confirmed with $F(z)=F_3(z)$. Interestingly, the
only parameter relevant asymptotically is the overall length input
rate $\lambda$.

The limiting behaviors of the scaling distribution are
\begin{equation}
\label{fz}
F(z)\simeq 
\cases{{1\over 3}-{1\over 4}z & $z\ll 1$,\cr
2z^{-3}                       &$z\gg 1$.}
\end{equation}
In particular, the large $z$ behavior implies the correct asymptotic
$P_\infty(x)\sim x^{-3}$, in agreement with Eq.~(\ref{alg}).  Thus, for
sufficiently large fragments, $x\gg t^{-1}$, the distribution has
already reached the final limiting form, while smaller sizes are still
created.

The formal solution (\ref{full}) has an interesting ``staircase''
structure, a time power series whose terms are products of time
independent functions $f_k(x)$ and time dependent functions $F_k(xt)$.
In fact, the solution for the random scission model in the absence of
input is also characterized by a similar structure. Indeed, consider
the evolution equation
\begin{equation}
\label{rate1} 
{\partial \tilde P(x,t)\over \partial t}=
-x\tilde P(x,t)+2\int_x^\infty dy\,\tilde P(y,t)
\end{equation} 
corresponding to the above fragmentation process in the absence of
input.  Given the initial conditions \hbox{$\tilde P(x,0)=\tilde
  f(x)$}, the solution can be obtained following the same steps that
led to Eq.~(\ref{full}).  Again, the full time dependent solution is 
a three term expansion: 
\begin{equation}
\label{pxt1}
\tilde P(x,t)=\sum_{k=1}^3 t^{k-1} \tilde f_k(x)\,\tilde F_k(xt).
\end{equation} 
The time independent functions are given by the same expressions
(\ref{fk}) as in the input case (with $f_k$ replaced by $\tilde f_k$),
while the time dependent functions are different from (\ref{Fk}):
\hbox{$\tilde F_1(z)=\tilde F_3(z)=e^{-z}$}, and \hbox{$\tilde
  F_2(z)=(2-z)e^{-z}$}.  In the limit $t\to\infty$, $x\to 0$ with
$z=xt$ kept finite, the scaling behavior emerges again.  Specifically,
\hbox{$\tilde P(x,t)\simeq \tilde\lambda t^2\tilde F(z)$} with
\hbox{$\tilde\lambda=\int dx\, x\tilde f(x)$} and the exponential
scaling function \hbox{$\tilde F(z)=e^{-z}$}.

The above solutions for the input case with empty initial conditions
and no input case with arbitrary initial conditions can be used to
construct the general solution for Eq.~(\ref{rate}).  Indeed, the sum
of the solutions (\ref{full}) and (\ref{pxt1}), $P(x,t)+\tilde
P(x,t)$, is the solution for a fragmentation process with input $f(x)$
starting from an initial distribution $\tilde f(x)$.  As expected, the
initial conditions are ``forgotten'' in the long time limit as
$P(x,t)$ given by Eq.~(\ref{full}) dominates over $\tilde P(x,t)$
given by Eq.~(\ref{pxt1}).  In particular, the scaling solution
(\ref{sclfrm}) is recovered, and the $P_{\infty}(x)\simeq 2\lambda
x^{-3}$ divergence of the limiting distribution holds in general. 

To examine the robustness of the algebraic behavior above, we consider
a natural generalization to $d$ spatial dimensions
\cite{tv,kb,rh,btv}.  Given that the most interesting long time
behavior is independent of the details of the source term, we focus on
the simplest monodisperse inputs, namely, unit hypercubes.  For
instance, in two dimensions we add unit squares with rate $\lambda$.
A unit square is divided by choosing a point $(x_1, x_2)$ with a
uniform probability density, and cutting the original square into four
rectangles of sizes $x_1 \times x_2$, $x_1 \times (1-x_2)$, $(1-x_1)
\times x_2$, and $(1-x_1) \times (1-x_2)$.  Similarly, the process is
repeated with rectangular fragments.

Let $P({\bf x},t)$ with ${\bf x}\equiv (x_1,\ldots,x_d)$ be the
distribution of fragments of size $x_1\times \cdots \times x_d$ at
time $t$. This quantity evolves according to the rate equation 
\begin{equation}
\label{rated}
\left({\partial \over \partial t}+|{\bf x}|\right) P({\bf x},t) 
=2^d\int_{\bf x} d{\bf y}\,P({\bf y},t)
+\lambda\delta({\bf x}-{\bf 1}).
\end{equation} 
Here, we used the shorthand notations ${\bf 1}=(1,\ldots,1)$ and $|{\bf
x}|=x_1\cdots x_d$.  The $d$-dimensional Mellin transform, $M({\bf
s},t)= \int d{\bf x}\,x_1^{s_1-1}\cdots x_d^{s_d-1} P({\bf x},t)$ with
${\bf s}\equiv (s_1,\ldots,s_d)$, reduces Eq.~(\ref{rated}) to
\begin{equation}
\label{msdt}
{\partial M({\bf s},t)\over \partial t}=
\left({2^d-s_1\cdots s_d\over s_1\cdots s_d}\right)
M({\bf s}+{\bf 1},t)+\lambda.
\end{equation}

We focus on the limiting size distribution $P_{\infty}({\bf x})$.  Its
Mellin transform $M_{\infty}({\bf s})$ is found from Eq.~(\ref{msdt})
by setting the time derivative to zero.  One gets
\hbox{$M_{\infty}({\bf s})=\lambda\left(1+{2^d\over (s_1-1)\cdots
      (s_d-1)-2^d}\right)$}.  Inverting this relation yields
\cite{kgb}
\begin{equation}
\label{pinfsol} 
P_\infty({\bf x})=\lambda\left[\delta({\bf x}-{\bf 1})
+2^d |{\bf x}|^{-1}\Phi_d(\xi)\right],
\end{equation}
with the shorthand notations
\begin{eqnarray}
\label{fd} 
\Phi_d(\xi)=\sum_{n=0}^\infty \left(\xi^n\over n!\right)^d 
\quad {\rm and} \quad 
\xi= 2\left(\prod_{i=1}^d \ln{1\over x_i}\right)^{1/d}.
\end{eqnarray}
In one dimension, $\Phi_1(\xi)=e^\xi=x^{-2}$, and we recover the
one-dimensional result $P_\infty(x)=2\lambda x^{-3}$.  In two
dimensions, $\Phi_2(\xi)=I_0(2\xi)$ where $I_0$ is the modified Bessel
function, and in general $\Phi_d(\xi)$ can be expressed in terms of
hypergeometric functions.

The small size behavior of $P_\infty({\bf x})$ can be obtained by using the
steepest decent method. The leading tail behavior, \hbox{$\Phi_d(\xi)\simeq
  (2\pi \xi)^{1-d\over 2} e^{\xi d}$} for $\xi\gg 1$, corresponds to the case
when at least one of the lengths is small, i.e., $x_i\ll 1$.  Returning
to the original variables, we re-write the above asymptotic as 
\begin{eqnarray*}
P_\infty({\bf x})\sim |{\bf x}|^{-1}|\ln {\bf x}|^{-{d-1\over 2d}} 
\exp\left[2d\left(|\ln {\bf x}|\right)^{1/d}\right],
\end{eqnarray*}
where $|\ln {\bf x}|\equiv \prod_{i=1}^d \ln {1\over x_i}$.  Thus, the
fragment distribution exhibits a ``log-stretched-exponential''
behavior.

Let us consider the limiting volume distribution $P_\infty(V)$ defined
via $P_\infty(V)=\int d{\bf x}\, P_\infty({\bf x})\,
\delta\left(V-x_1\cdots x_d\right)$.  Its Mellin transform,
\hbox{$M_{\infty}(s)=\int dV\,V^{s-1} P_\infty (V)$}, immediately
follows from Eq.~(\ref{msdt}):
\hbox{$M_{\infty}(s)=\lambda\left[1+{2^d\over (s-1)^d-2^d}\right]$}.
Using the identity $(a^d-1)^{-1}=d^{-1}\sum_{k=0}^{d-1}
\zeta^k(a-\zeta^k)^{-1}$, where $\zeta=e^{2\pi i/d}$ is the primitive
$d^{\rm th}$ root of unity, we can express $M_{\infty}(s)$ as a sum
over simple poles at $1+2\zeta^k$.  Consequently, the inverse Mellin
transform is given by a linear combination of $d$ power laws,
\begin{equation}
\label{pv}
P_\infty(V)=\lambda\left[\delta(V-1)+{2\over d}
\sum_{k=0}^{d-1} \zeta^k V^{-1-2\zeta^k}\right].
\end{equation}
One can verify that the volume distribution is real since it equals
its complex conjugate. The small-volume tail of the distribution can
be obtained by noting that the sum in Eq.~(\ref{pv}) is dominated by
the first term in the series, 
\begin{equation}
\label{pvalg} 
P_\infty(V)\simeq {2\lambda\over d}\, V^{-3}, \quad{\rm for}\quad V\to 0.
\end{equation}
Thus, the same $V^{-3}$ algebraic behavior occurs in all spatial
dimensions.  Clearly, this divergence is general. Indeed, the time
dependent evolution equation (\ref{rated}) implies that the overall
volume grows linearly, and that the overall number of fragments grows
quadratically.  Therefore, the heuristic scaling argument leading to
Eq.~(\ref{sclfrm}) extends to higher dimensions, and consequently, the
limiting behavior is given by Eq.~(\ref{pvalg}).

In summary, we have studied random fragmentation processes in the
presence of a steady source. We have solved for the full time
dependent behavior in terms of the input function. In the long time
limit, the size distribution exhibits a universal scaling behavior.
The limiting distribution diverges algebraically according to $x^{-3}$
in the small size limit. This behavior is robust. It applies to
arbitrary inputs, and it extends to higher dimensions as well.
Interestingly, the only asymptotically relevant parameter is the total
volume added per unit time. Additionally, we have shown that the
scaling behavior can be used to predict the algebraic nature of the
final size distribution. Hence, the scaling behavior and the limiting
distribution are closely related.

The solution for the time dependent behavior exhibits an interesting
staircase structure. The two progressively weaker corrections to the
leading behavior are of the order $t^{-1}$ and $t^{-2}$, respectively.
Such staircase structures may be a useful tool for treating similar
integro-differential equations which are expected to exhibit a scaling
asymptotic behavior.  For instance, one can check whether substituting
such an ansatz leads to a closed system of equations for the time
dependent and time independent functions.

The above results can be extended in a number of ways.  One may try to
derive a general solution for $P({\bf x},t)$ in higher dimensions for
arbitrary input rate $f({\bf x})$.  We anticipate that geometric
features of the fragments will be interesting.  Indeed, in the no
input case the volume distribution exhibits an ordinary scaling
behavior while multiscaling asymptotic behavior underlies the full
multivariate size distribution.  Additionally, a nontrivial set of
conservation laws exists as all moments $M({\bf s}^*,t)$ with
$\prod_{i=1}^d s_i^*=2^d$ are conserved, as seen from
Eq.~(\ref{msdt}).  In the presence of input, the same moments grow
linearly in time but multiscaling should still hold asymptotically.

\medskip\noindent
We acknowledge support from DOE (W-7405-ENG-36),
NSF (DMR9978902), ARO (DAAD19-99-1-0173), and CNRS.

\end{multicols}
\end{document}